\documentclass[journal=jacsat,manuscript=communication,articletitle=true,layout=twocolumn,]{achemso}
\usepackage[usenames,dvipsnames]{xcolor}
\usepackage[bookmarks=false,colorlinks]{hyperref}
\hypersetup{
	linkcolor=blue, 
	citecolor=blue, 
	filecolor=black, 
	urlcolor=black,  
	urlcolor=blue
} 
\usepackage{amsmath,amssymb}
\usepackage{mdwlist} 
\usepackage{booktabs}
\usepackage{multirow}
\usepackage{amsmath}
\usepackage{adjustbox}
\usepackage{float}
\usepackage{mhchem}
\usepackage[T1]{fontenc}

\definecolor{col1}{rgb}{0.0, 0.46, 0.8}
\definecolor{col2}{rgb}{0.9, 0.0, 0.30}
\definecolor{col3}{rgb}{0,0,154}
\definecolor{darkblue}{rgb}{0.0, 0.0, 0.55}

\usepackage{lineno}

\title{Synergetic Enhancement of Power Factors and Suppression of Lattice Thermal Conductivities via Biaxial Strain in ScAgSe$_2$ and TmAgTe$_2$}

\author{Wu Xiong}
\affiliation{Key Laboratory of Advanced Materials and Devices for Post-Moore Chips, Ministry of Education, University of Science and Technology Beijing, Beijing 100083, China}
\alsoaffiliation{School of Mathematics and Physics, University of Science and Technology Beijing, Beijing 100083, China}

\author{Zhongjuan Han}
\affiliation{Key Laboratory of Advanced Materials and Devices for Post-Moore Chips, Ministry of Education, University of Science and Technology Beijing, Beijing 100083, China}
\alsoaffiliation{School of Mathematics and Physics, University of Science and Technology Beijing, Beijing 100083, China}

\author{Zhonghao Xia}
\affiliation{Key Laboratory of Advanced Materials and Devices for Post-Moore Chips, Ministry of Education, University of Science and Technology Beijing, Beijing 100083, China}
\alsoaffiliation{School of Mathematics and Physics, University of Science and Technology Beijing, Beijing 100083, China}

\author{Zhilong Yang}
\affiliation{Key Laboratory of Advanced Materials and Devices for Post-Moore Chips, Ministry of Education, University of Science and Technology Beijing, Beijing 100083, China}
\alsoaffiliation{School of Mathematics and Physics, University of Science and Technology Beijing, Beijing 100083, China}

\author{Jiangang He}
\email{jghe2021@ustb.edu.cn}
\affiliation{Key Laboratory of Advanced Materials and Devices for Post-Moore Chips, Ministry of Education, University of Science and Technology Beijing, Beijing 100083, China}
\alsoaffiliation{School of Mathematics and Physics, University of Science and Technology Beijing, Beijing 100083, China}

\begin{document}
	\begin{abstract}
	\noindent The challenge of achieving high thermoelectric (TE) performance is mainly from the entanglement among Seebeck coefficient ($S$), electrical conductivity ($\sigma$), and lattice thermal conductivity ($\kappa_{\mathrm{L}}$). In this work, we propose a synergetic strategy of enhancing power factor (PF, $S^2\sigma$) and suppressing $\kappa_{\mathrm{L}}$ by applying a biaxial tensile strain in two silver chalcogenides ScAgSe$_2$ and TmAgTe$_2$ with TlCdS$_2$-type structure. The forbidden $p$-$d$ orbital coupling at the $\Gamma$ point and allowed $p$-$d$ orbital coupling at the A point and the middle of $\Lambda$ line leads to high electronic band dispersion along the $\Gamma$-A direction and a high-degeneracy valence band valley ($\Lambda_2$). The elongation of the Ag-Se bond under tensile strain weakens the orbital coupling between Ag-$d$ and Se/Te-$p$ orbitals and reduces the band energy at the A point, which aligns the valence band and achieving a high band degeneracy. Concurrently, the weaker Ag-Se/Ag-Te bond under a small tensile strain leads to lower phonon group velocity and strong three- and four phonon scatterings, leading to lower $\kappa_{\mathrm{L}}$. Our first-principles calculations combined with electron-phonon coupling analysis as well as phonon and electron Boltzmann transport equations show that applying a 3\% (2\%) tensile strain can enhance the PF along the $c$-axis of ScAgSe$_2$ (TmAgTe$_2$) by 243\% (246\%) at a carrier concentration of 3$\times$10$^{20}$ cm$^{-3}$ and reduce the $\kappa_{\mathrm{L}}$ by 37\% (26\%) at 300 K. Consequently, 2 $\sim$ 4 times of $ZT$ enhancement is obtained by 3\% or 1\% tensile strain in ScAgSe$_2$ (TmAgTe$_2$) at 300 K, achieving a maximum $ZT$ of 3.10 (3.62) at 800 K. Our material design strategy based on molecular orbital analysis reveals an effective route to boosting TE performance, and can be extended to other systems as well.
	
	\end{abstract}

	\maketitle
	
	\noindent $\blacksquare$ \textbf{\textcolor{darkblue}{INTRODUCTION}} \\
	TE technology, which enables the direct interconversion of thermal and electrical energy, has emerged as a promising route for sustainable-energy applications, particularly in waste-heat recovery and solid-state refrigeration~\cite{science.1159725,2022Chemical}. The performance of a TE material is typically quantified by the dimensionless figure of merit $ZT = S^{2}\sigma T/(\kappa_{\mathrm{L}}+\kappa_{\mathrm{e}})$, where $S$, $\sigma$, and $T$ denote the Seebeck coefficient, electrical conductivity, and absolute temperature, respectively, and $\kappa_{\mathrm{L}}$ and $\kappa_{\mathrm{e}}$ are the lattice and electronic contributions to the thermal conductivity. However, the inherent interdependence among $S$, $\sigma$, and $\kappa_{\mathrm{L}}$ fundamentally constrains the improvement of $ZT$. For instance, low-carrier-density insulators and semiconductors generally exhibit large $S$ but low $\sigma$, whereas heavily doped semiconductors and metals show high $\sigma$ but low $S$~\cite{2008Complex}. In addition, semiconductors with low $\kappa_{\mathrm{L}}$ often also display low $\sigma$ because, within the acoustic deformation–potential scattering regime, $\sigma$ scales with elastic stiffness (e.g., the bulk modulus)~\cite{PhysRev.80.72}, while a large bulk modulus typically reflects strong bonding interactions, which in turn lead to high $\kappa_{\mathrm{L}}$~\cite{https://doi.org/10.1002/anie.201508381}.

	Band alignment (valley convergence) is an effective strategy to enhance the PF of TE materials~\cite{moshwan2019realizing,long2023band}. Within the single-parabolic-band (SPB) model~\cite{2018A}, $S$ scales with the density-of-states effective mass ($m_{\mathrm{d}}^{*}$), which depends on both the band effective mass ($m_{\mathrm{b}}^{*}$) and the band degeneracy ($N_{\mathrm{v}}$). The latter can be expressed as $N_{\mathrm{v}} = N_{\mathrm{vk}} \times N_{\mathrm{vo}}$, where $N_{\mathrm{vk}}$ is the $k$-space valley multiplicity and $N_{\mathrm{vo}}$ is the orbital degeneracy. Increasing $N_{\mathrm{vk}}$ can elevate $m_{\mathrm{d}}^{*}$ without proportionally penalizing the carrier mobility ($\mu$), thereby mitigating the conventional trade-off between $S$ and $\sigma$ and boosting the PF~\cite{2012Convergence,doi:10.1021/jacs.4c04048}. According to the Wiedemann–Franz law~\cite{solidstatephysics}, $\kappa_{\mathrm{e}}$ is proportional to $\sigma$ at a given T. Consequently, improving $ZT$ requires suppressing $\kappa_{\mathrm{L}}$ while maximizing the PF. Within a kinetic theory~\cite{tritt2005thermal}, $\kappa_{\mathrm{L}} = \tfrac{1}{3} C_{v}\, \nu_{\mathrm{g}}^{2}\, \tau$,
	where $C_{v}$ is the volumetric heat capacity, $\nu_{\mathrm{g}}$ is the average phonon group velocity, and $\tau$ is the phonon relaxation time. The group velocity scales as $\nu_{\mathrm{g}} \propto \sqrt{k/M}$, where $k$ is an effective bond stiffness (interatomic force constant) and $M$ is the average atomic mass. Hence, materials with shorter $\tau$ and weaker bonding (small $k$), and larger $M$ generally exhibit lower $\kappa_{\mathrm{L}}$. Established approaches to shorten $\tau$ and lower $\nu_{\mathrm{g}}$ include introducing point defects~\cite{Mao03042018}, nanoscale precipitates~\cite{doi:10.1126/science.1092963,doi:10.1126/science.1156446}, stereochemically active lone pairs (electrons/ions)~\cite{PhysRevLett.107.235901,2013Lone}, rattling phonon modes~\cite{2015Impact,2016Ultralow}, and intrinsically weak chemical bonds~\cite{https://doi.org/10.1002/adfm.202108532,https://doi.org/10.1002/advs.202417292,he2019designing}. Nevertheless, concurrent optimization of the PF and $\kappa_{\mathrm{L}}$ remains challenging and is rarely achieved in practice.

	Additionally, epitaxial strain or hydrostatic pressure can be exploited to tailor the TE properties of materials~\cite{10.1063/1.4866861}. For example, previous studies have shown that compressive strain (or hydrostatic pressure) can increase valley degeneracy in IV–VI TE semiconductors, thereby substantially enhancing the PF~\cite{10.1063/1.4866861,dai2022simultaneous}. However, compressive strain generally strengthens bonding interactions and increases $\nu_{\mathrm{g}}$, which elevates $\kappa_{\mathrm{L}}$, whereas tensile strain typically reduces $\kappa_{\mathrm{L}}$ by weakening chemical bonds, softening phonons, and lowering $\nu_{\mathrm{g}}$~\cite{gupta2023enhancement}. Consequently, achieving synergistic optimization of electronic and phononic transport—simultaneously maximizing the PF while minimizing $\kappa_{\mathrm{L}}$—within a single material remains challenging and is rarely realized.

	In this work, we apply biaxial tensile strain within the $ab$ plane to the TlCdS$_2$-type ScAgSe$_2$ and TmAgTe$_2$, which have been previously identified as promising thermoelectrics~\cite{sc,tm}, to optimize their TE performance. Our calculations show that a small tensile strain reduces the Ag-$d_{x^2-y^2}$–Se-$p_{x}/p_{y}$ hybridization and lowers the energy of the valence-band maximum (VBM), bringing it into near-degeneracy with the second-highest valence band. The increased band degeneracy (from 1 to 7) enhances the PF by a factor of 2--3 at 300~K. Simultaneously, strain weakens the Ag-Se/Ag-Te and Sc-Se/Tm-Te bonds, decreases the average phonon group velocity ($\overline{\nu_\mathrm{g}}$), and thereby suppresses $\kappa_{\mathrm{L}}$. Under modest (1\%--3\%) tensile strain, the PF along the $c$ axis increases, for ScAgSe$_2$ (TmAgTe$_2$), from 4.57 (20.58) to 21.95 (29.16) $\mu\mathrm{W}\,\mathrm{cm}^{-1}\,\mathrm{K}^{-2}$ at 300~K and from 10.65 (12.39) to 15.11 (20.18) $\mu\mathrm{W}\,\mathrm{cm}^{-1}\,\mathrm{K}^{-2}$ at 800~K. Concurrently, $\kappa_{\mathrm{L}}$ along the $c$ axis decreases from 0.57 (0.38) $\mathrm{W}\,\mathrm{m}^{-1}\,\mathrm{K}^{-1}$ in the bulk to 0.36 (0.27) $\mathrm{W}\,\mathrm{m}^{-1}\,\mathrm{K}^{-1}$ at 300~K and 800~K, respectively. Consequently, the maximum $ZT$ along the $c$ axis reaches 1.23 (1.22) at 300~K and 3.10 (3.62) at 800~K, representing improvements of 412\% (259\%) and 130\% (126\%) relative to the unstrained counterparts. The substantial enhancement of $ZT$ at room and intermediate temperatures is particularly noteworthy, because the temperature-averaged PF and $ZT$ over the operating range are more relevant to conversion efficiency than peak values in practical thermoelectric applications.

	\vspace{0.5 cm}
	\noindent $\blacksquare$ \textbf{\textcolor{darkblue}{COMPUTATIONAL METHODS}} \\	
	The structural relaxation, transport effective mass, and lattice dynamics calculations were performed using the Vienna {\it ab initio} Simulation Package (VASP)~\cite{vasp1,vasp2}. All calculations employed the Perdew-Burke-Ernzerhof (PBE) exchange-correlation functional~\cite{PBE} with a plane-wave basis set and an energy cutoff of 520~eV. The second-order force constants were obtained using PHONOPY~\cite{phonopy}, while the third- and fourth-order constants were determined via compressive sensing lattice dynamics (CSLD)~\cite{CSLD}, based on VASP-calculated forces for a $4\times4\times2$ supercell (128 atoms) with a $2\times2\times3$ $k$-point mesh. Finite-temperature phonon properties were computed using self-consistent phonon (SCPH) theory~\cite{scph1,scph2,scph3}. The $\kappa_{\mathrm{L}}$ was evaluated by solving the Peierls-Boltzmann transport equation with a $24\times24\times12$ $q$-mesh for the HA+3ph ($\kappa^{\mathrm{HA}}_{\mathrm{3ph}}$) and SCPH+3ph ($\kappa^{\mathrm{SCPH}}_{\mathrm{3ph}}$) schemes, and a $20\times20\times10$ $q$-mesh for SCPH+3,4ph ($\kappa^{\mathrm{SCPH}}_{\mathrm{3,4ph}}$) calculations. For biaxial strain applications within the $ab$ plane, atomic positions and the lattice constant $c$ were fully relaxed while fixing $a$ and $b$. The biaxial strain, $\varepsilon$, is defined as $\varepsilon = (a - a_0)/a_0$, where $a$ and $a_0$ are the strained and equilibrium lattice constants $a$ ($b$), respectively. Crystal structures were visualized using VESTA~\cite{vesta}.
	
	The transport effective mass~\cite{gibbs2017effective}, defined as $m^{*}(T,\mu) = \frac{e^2 \tau}{\sigma(T,\mu)}\,n(T,\mu)$,
	where $e$ is the electron charge, $n$ is the carrier concentration, and $\mu$ is the chemical potential, was computed using BoltzTraP2~\cite{BoltzTraP2} based on VASP-calculated electronic structures. These calculations employed a self-consistent field (SCF) convergence threshold of $10^{-8}$~eV, reciprocal-space projection operators (LREAL = FALSE), and a dense $\Gamma$-centered $k$-point mesh for Brillouin zone sampling.
	
	The electronic structure calculations were carried out using the Quantum ESPRESSO (QE) package~\cite{QE} with the Perdew–Burke–Ernzerhof (PBE) exchange–correlation functional~\cite{PBE} and Rabe-Rappe-Kaxiras-Joannopoulos (RRKJ)~\cite{RRKJ} ultrasoft pseudopotential. A plane-wave kinetic energy cut-off of 80~Ry was employed. The SCF calculations were performed using a $12 \times 12 \times 6$ $k$-point mesh. The electronic band structure was interpolated using maximally localized Wannier functions (MLWFs) generated with Wannier90~\cite{wannier90}. Phonon properties obtained from QE were combined with MLWF-based electronic states in PERTURBO~\cite{Perturbo} to evaluate phonon-limited electron transport. The transport properties were computed by solving the Boltzmann transport equation (BTE) within the relaxation time approximation (RTA).
	
	\begin{table*}
	\scriptsize
	\setlength{\tabcolsep}{1pt} 
	\renewcommand{\arraystretch}{1.3} 
	\centering
	\caption{Summary of calculated structural, electronic, phononic, and thermoelectric properties of ScAgSe$_2$ and TmAgTe$_2$ under different biaxial strains ($\varepsilon = (a - a_0)/a_0$). Listed parameters include the lattice constants ($a$, $c$), bond angle $\angle(X\mathrm{-Ag-}X)$, $-$iCOHP values for Ag–Se/Ag–Te, Ag–$X$ bond length ($l$), energy difference between the highest valence band at $\Lambda_1$ and VBM ($\Delta E$), and between the highest valence band at $\Sigma$ and the VBM ($\Delta E^\prime$), band gaps ($E_\mathrm{g}$), average phonon group velocities ($\overline{\nu_\mathrm{g}}$), maximum $ZT$ values in the out-of-plane direction at 300~K and 800~K (with the corresponding optimal hole concentrations $n_\mathrm{h}$ in cm$^{-3}$ given in parentheses), and the associated PF, $\kappa_{\mathrm{L}}$, and $ZT$ values.}	
	\begin{tabular}{cccccccccccccc}
		\hline
		\hline
		Comp.&$\varepsilon$&$a$  & $c$ &$\angle(X\mathrm{-Ag-}X)$& $l$ &-iCOHP&$\Delta{E}$&$\Delta{E}^\prime$&E$_\mathrm{g}$&$\overline{\nu_\mathrm{g}}$& PF 300K,800K&$\kappa_{\mathrm{L}}$ 300K,800K& $ZT_\mathrm{max}$ 300K,800K\\
     	&    \%       &(\AA)&(\AA)&      ($^{\circ}$)       &(\AA)&      &  (meV)  &      (meV)       &  (eV)        &        (m/s)            &($\mu$Wcm$^{-1}$K$^{-2}$)&(Wm$^{-1}$K$^{-1}$)&  \\
		\hline
		           & 0\% & 3.9234 & 6.6267 & 86.04 & 2.8754 & 0.66230 & 205 & 316 & 0.67 & 2193 &  4.57,10.65 & 0.57,0.38 & 0.24(2$\times$10$^{19}$),1.35(8$\times$10$^{19}$) \\
		ScAgSe$_2$ & 2\% & 4.0019 & 6.5069 & 87.53 & 2.8927 & 0.64961 &  89 & 171 & 0.80 & 2100 & 21.32,18.45 & 0.47,0.34 & 0.68(2$\times$10$^{20}$),2.48(2$\times$10$^{20}$) \\
		           & 3\% & 4.0411 & 6.4523 & 88.22 & 2.9029 & 0.64147 &  35 & 105 & 0.87 & 2034 & 21.95,15.11 & 0.36,0.27 & 1.23(8$\times$10$^{19}$),3.10(1$\times$10$^{20}$) \\
		\hline
	               & 0\% & 4.3390 & 6.9825 & 91.06 & 3.0402 & 0.49235 & 165 & 246 & 0.66 & 1823 & 20.58,12.39 & 0.62,0.42 & 0.34(4$\times$10$^{20}$),1.60(1$\times$10$^{20}$) \\
		TmAgTe$_2$ & 1\% & 4.3824 & 6.9207 & 91.68 & 3.0545 & 0.47367 &  93 & 156 & 0.78 & 1728 & 21.91,15.72 & 0.39,0.24 & 0.79(1$\times$10$^{20}$),3.62(5$\times$10$^{19}$) \\
		           & 2\% & 4.4258 & 6.8643 & 92.27 & 3.0695 & 0.46497 &  23 &  71 & 0.87 & 1656 & 29.16,20.18 & 0.46,0.32 & 1.22(7$\times$10$^{19}$),3.40(7$\times$10$^{19}$) \\
		\hline
	\end{tabular}
	\label{lattice constant}
	\end{table*}  

	\begin{table}
		\scriptsize
		\setlength{\tabcolsep}{2.5pt} 
		\renewcommand{\arraystretch}{1.3} 
		\centering
		\caption{Atomic valence-electron band representations (ABRs) of ScAgSe$_2$ ($P\overline{3}m1$) and the corresponding elementary band representations (EBRs) they induce at various high-symmetry points in the Brillouin zone.}
		\begin{tabular}{ccccc}
			\hline
			\hline
			Orbitrals & \multicolumn{2}{c}{Ag $d$} & \multicolumn{2}{c}{Se $p$} \\
						\cmidrule(l{6pt}r{6pt}){1-1} \cmidrule(l{6pt}r{6pt}){2-3} \cmidrule(l{6pt}r{6pt}){4-5}
			WKS ($q$) & \multicolumn{2}{c}{1$b$}   & \multicolumn{2}{c}{2$d$}   \\
			Irrep. ($p$)     & A$_{\mathrm{1g}}$      & E$_{\mathrm{g}}$      & A$_1$      & E      \\
			Basis     & $d_{z^2}$ &($d_{x^2-y^2}$,$d_{xy}$),  & $p_z$ & ($p_x$,$p_y$) \\
			          &      &    ($d_{xz}$,$d_{yz}$) &       &       \\
			ABRs ($p@q$)     & A$_{\mathrm{1g}}$@1$b$ & E$_{\mathrm{g}}$@1$b$ & A$_1$@2$d$ & E@2$d$ \\
			EBRs ($\Lambda$) & $\Lambda_1$  &  $\Lambda_1\oplus\Lambda_2$ & $\Lambda_1$$\oplus$$\Lambda_2$   & $\Lambda_1$$\oplus$$\Lambda_2$   \\
			EBRs ($\Gamma$)  & $\Gamma_1^+$ & $\Gamma_3^+$ & $\Gamma_1^+$$\oplus$$\Gamma_2^-$ & $\Gamma_3^+$$\oplus$$\Gamma_3^-$ \\
			EBRs (A)         & A$_2^-$ & A$_3^-$                & A$_1^+$$\oplus$A$_2^-$ & A$_3^+$$\oplus$A$_3^-$ \\
			EBRs (L)         & L$_2^-$ & L$_1^-$$\oplus$L$_2^-$ & L$_1^+$$\oplus$L$_2^-$ & L$_1^+$$\oplus$L$_1^-$$\oplus$L$_2^+$$\oplus$L$_2^-$ \\
			EBRs ($\Sigma$ ) & $\Sigma_1$  & $\Sigma_1\oplus\Sigma_2$ & $\Sigma_1$ & $\Sigma_1\oplus\Sigma_2$ \\
			\hline	
		\end{tabular}
		\label{EBR}
	\end{table}  

	\begin{figure*}[th!]
		\centering
		\includegraphics[width=0.9\linewidth]{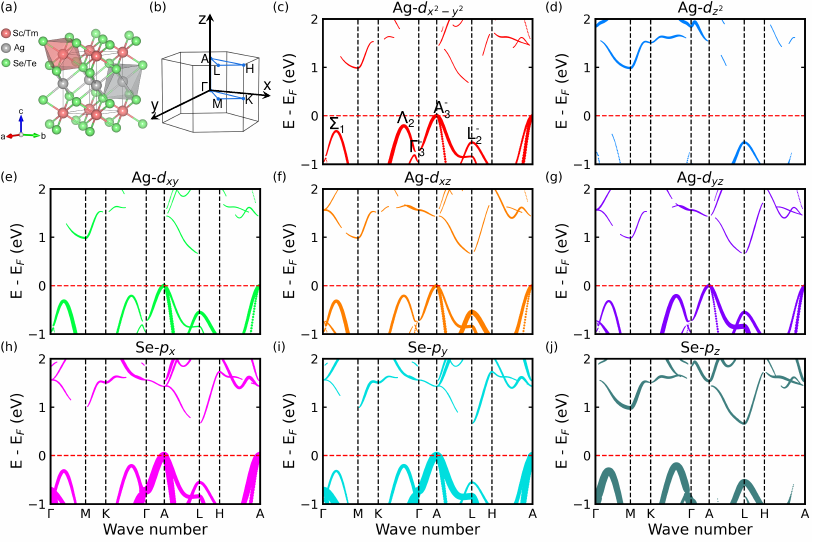}
		\caption{(a) and (b) Crystal structure and Brillouin zone of ScAgSe$_2$ (TmAgTe$_2$), respectively.  
			(c)–(j) Orbital-resolved electronic structures, showing the contributions of the five Ag $d$-orbitals ($d_{x^2-y^2}$, $d_{z^2}$, $d_{xy}$, $d_{xz}$, and $d_{yz}$) and the three Se $p$-orbitals ($p_x$, $p_y$, and $p_z$) to the band structure. The size of the dots indicates the magnitude of the orbital contribution. Panel (c) also labels the IRREPs of bands at the VBM located along the $\Lambda$ line, and at the $\Gamma$, A, and L points.}
		\label{band}
	\end{figure*}
	
	\begin{figure}[th!]
		\setlength{\unitlength}{1cm}
		\includegraphics[width=1.0\linewidth]{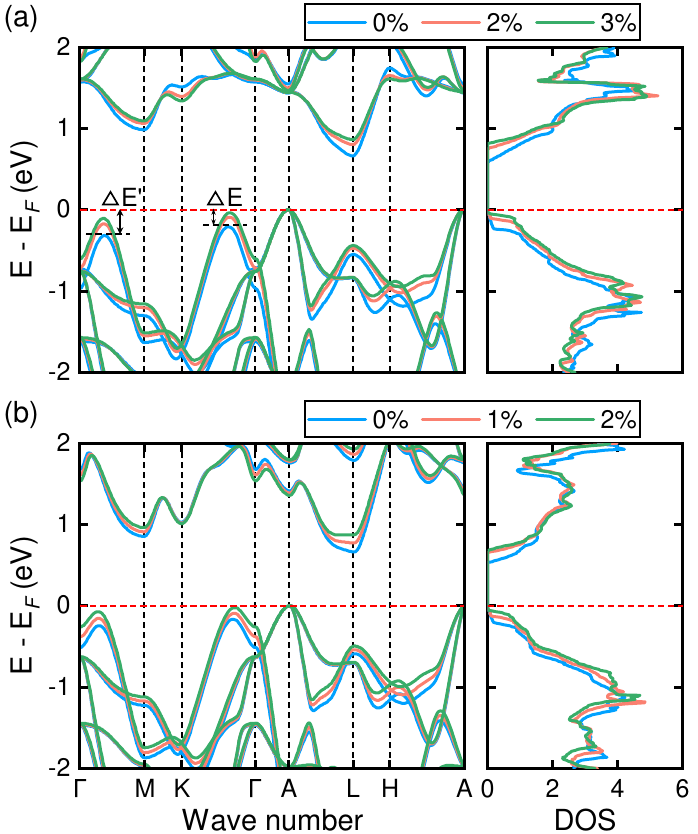}
		\caption{(a) and (b) are the electronic band structures and density of states (DOS) of ScAgSe$_2$ and TmAgTe$_2$ under different strains, respectively.}
		\label{bandstrain}
	\end{figure}

	\begin{figure*}[th!]
		\setlength{\unitlength}{1cm}
		\includegraphics[width=1.0\linewidth]{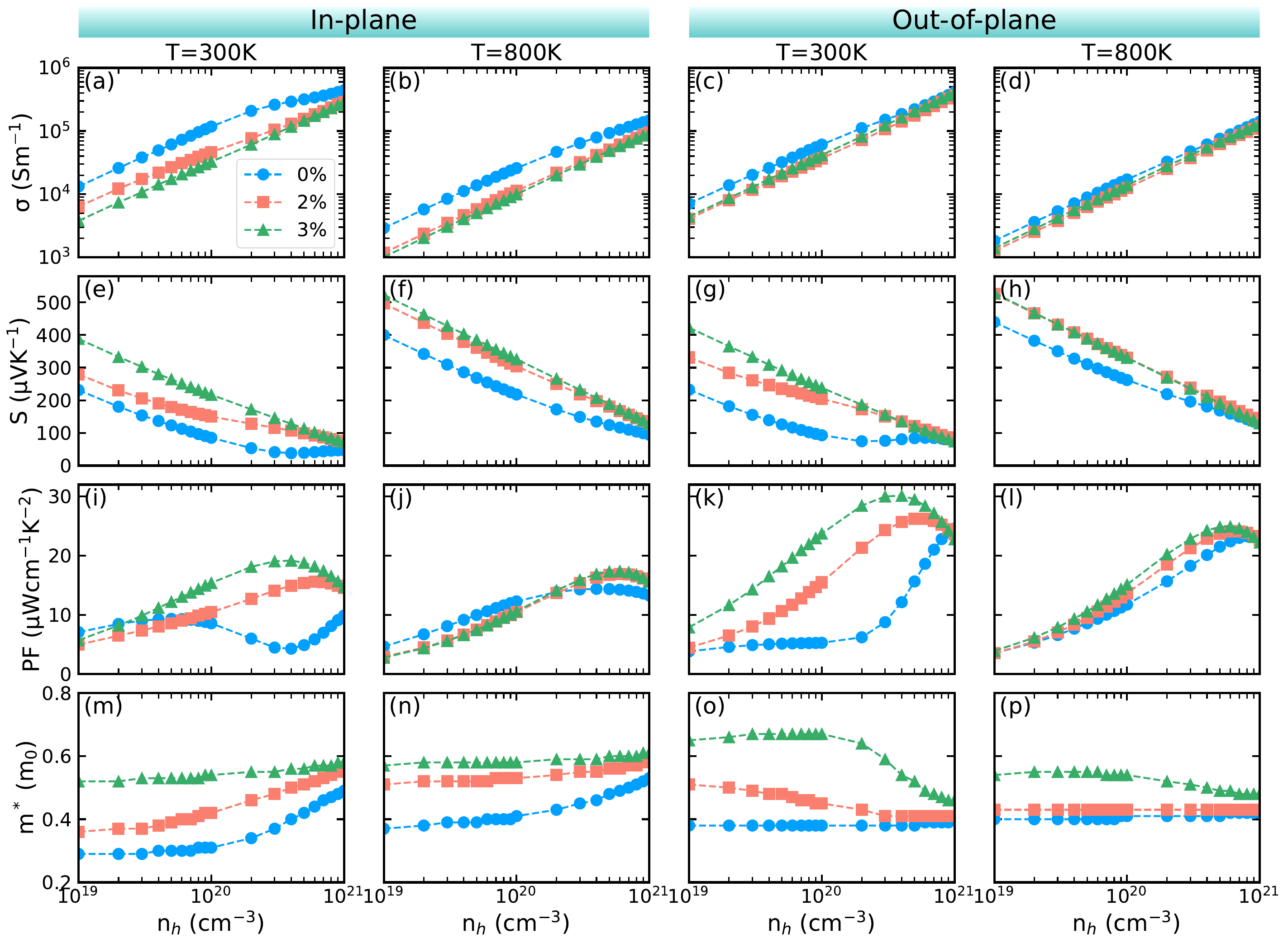}
		\caption{Variation of in-plane and out-of-plane electrical conductivity ($\sigma$), Seebeck coefficient ($S$), power factor (PF), and transport effective mass ($m^*$) of ScAgSe$_2$ as functions of hole concentration ($n_\mathrm{h}$) and biaxial strain magnitude at 300~K and 800~K. (a), (e), (i), and (m) show the in-plane $\sigma$, $S$, PF, and $m^*$ at 300~K, respectively. (b), (f), (j), and (n) show the in-plane $\sigma$, $S$, PF, and $m^*$ at 800~K, respectively. (c), (g), (k), and (o) show the out-of-plane $\sigma$, $S$, PF, and $m^*$ at 300~K, respectively. (d), (h), (l), and (p) show the out-of-plane $\sigma$, $S$, PF, and $m^*$ at 800~K, respectively.}
		\label{sigma}
	\end{figure*}

	\begin{figure*}[th!]
		\setlength{\unitlength}{1cm}
		\includegraphics[width=1.0\linewidth]{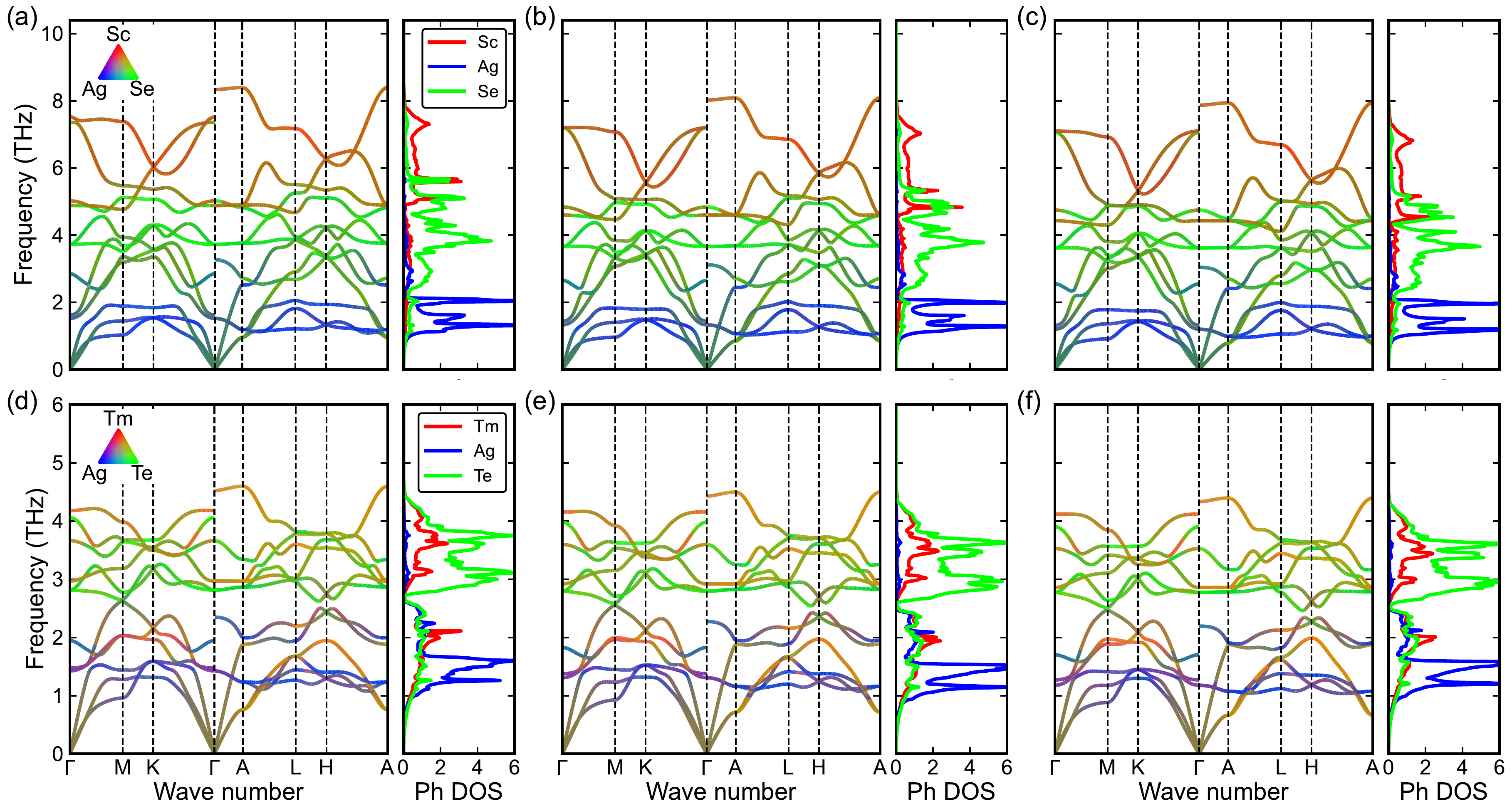}
		\caption{Phonon dispersion relations and phonon density of states (PhDOS) of ScAgSe$_2$ and TmAgTe$_2$ under different biaxial strains at 300~K. (a)–(c) show results for ScAgSe$_2$ under 0\%, 2\%, and 3\% biaxial strain, respectively. (d)–(f) show results for TmAgTe$_2$ under 0\%, 1\%, and 2\% biaxial strain, respectively.}
		\label{phonon}
	\end{figure*}

	\begin{figure}[th!]
		\centering
		\includegraphics[width=1.0\linewidth]{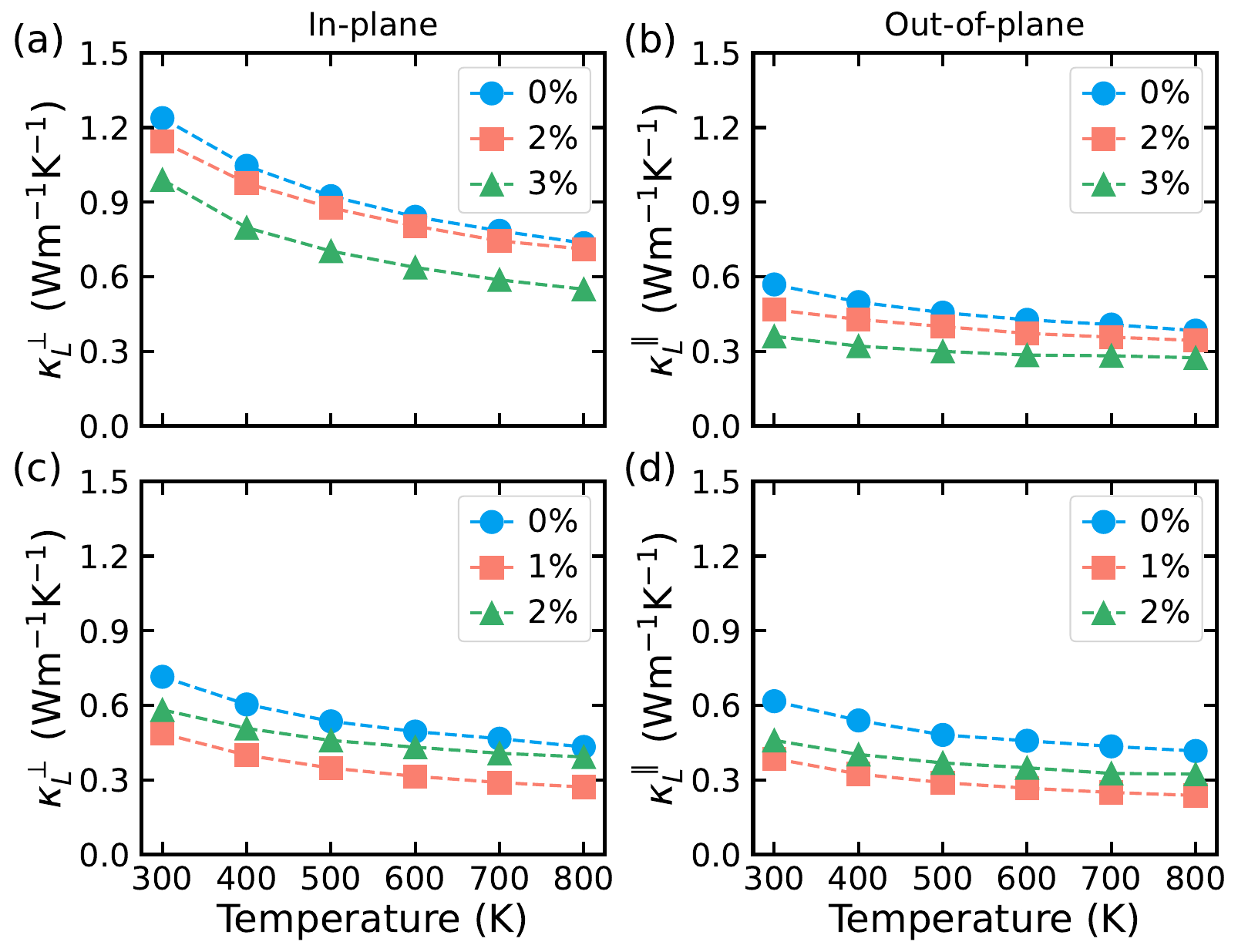}
		\caption{Temperature- and strain-dependent $\kappa_{\mathrm{L}}^{\perp}$ and $\kappa_{\mathrm{L}}^{\parallel}$ of ScAgSe$_2$ and TmAgTe$_2$. (a) and (b) show the $\kappa_{\mathrm{L}}^{\perp}$ and $\kappa_{\mathrm{L}}^{\parallel}$ of ScAgSe$_2$, respectively. (c) and (d) show the $\kappa_{\mathrm{L}}^{\perp}$ and $\kappa_{\mathrm{L}}^{\parallel}$ of TmAgTe$_2$, respectively.}
		\label{kappa}
	\end{figure}

	\begin{figure*}[th!]
		\setlength{\unitlength}{1cm}
		\includegraphics[width=1.0\linewidth]{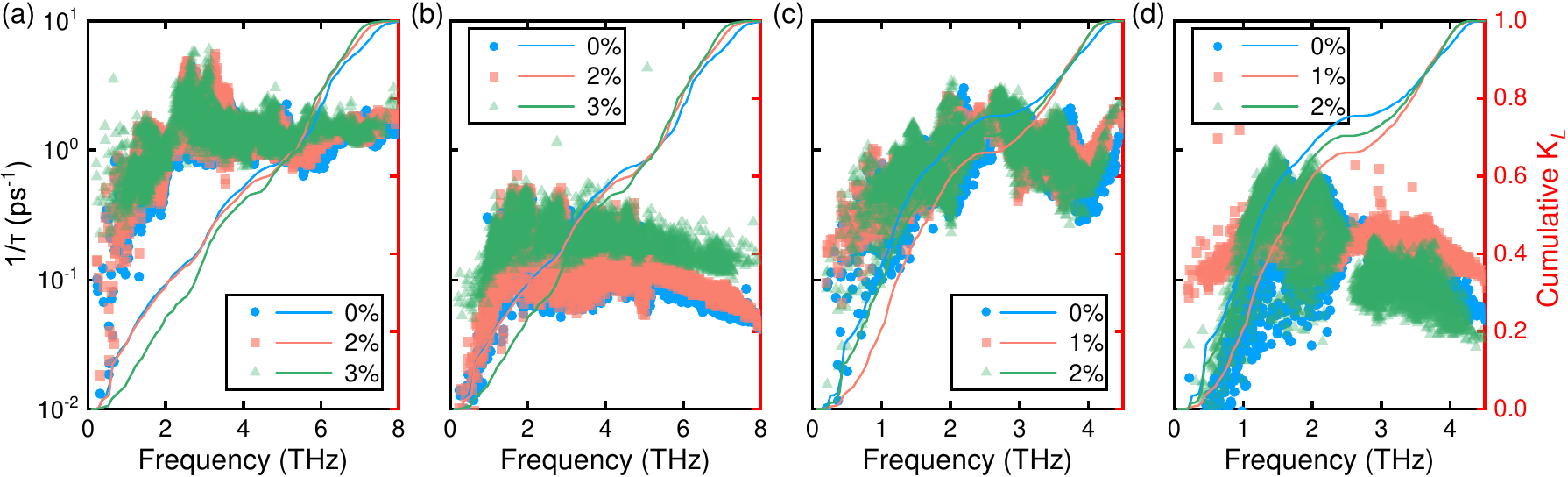}
		\caption{Three-phonon (3ph) and four-phonon (4ph) scattering rates ($1/\tau$) and cumulative lattice thermal conductivity of ScAgSe$_2$ and TmAgTe$_2$ as functions of phonon frequency and biaxial strain magnitude at 300 K. (a) and (b) show the 3ph and 4ph scattering rates of ScAgSe$_2$, respectively. (c) and (d) show the 3ph and 4ph scattering rates of TmAgTe$_2$, respectively.}
		\label{ss}
	\end{figure*}

	\vspace{0.5 cm}
	\noindent $\blacksquare$ \textbf{\textcolor{darkblue}{RESULTS AND DISCUSSION}} \\
	\textbf{Crystal structures.} The ternary compounds with the general formula $R$Ag$X_2$ ($R$ = Sc, Y, and rare-earth elements; $X$ = Se, Te) exhibit multiple polymorphic modifications~\cite{SACHANYUK20071091}. $R$AgSe$_2$ ($R$ = Sc, Y, Er, Ho, Dy, Tm, Yb, and Lu) crystallizes in the ErAgSe$_2$-type structure ($P2_12_12_1$), whereas $R$AgSe$_2$ ($R$ = Gd, Tb, Ho, and Yb) has also been reported to adopt the YbAgS$_2$-type structure ($I4_1md$). Similarly, $R$AgTe$_2$ ($R$ = Y, Gd, Dy, Er, Ho, Lu, and Tm) can crystallize in either the Er$_{0.66}$Cu$_2$S$_2$-type ($P\bar{3}$) or the ErAgTe$_2$-type ($P\bar{4}2_1m$) structures~\cite{GULAY2006159}. ScAgSe$_2$ and TmAgSe$_2$ with the TlCdS$_2$-type structure ($P\bar{3}m1$) were first synthesized by Shemet \textit{et al.}~\cite{SHEMET2006186} in 2006 and Gulay \textit{et al.}~\cite{GULAY2007L1} in 2007, respectively. The crystal structure of TlCdS$_2$-type ScAgSe$_2$ (TmAgTe$_2$) is shown in Figure~\ref{band}(a), featuring an alternating stacking of ScSe$_6$ (TmTe$_6$) and AgSe$_6$ (AgTe$_6$) octahedra along the $c$ axis. Within each stacking sequence, adjacent ScSe$_6$ and AgSe$_6$ octahedra are face-sharing, while octahedra of the same type share edges within a layer. Cation–anion bond lengths within each octahedron are identical. However, the bond angles deviate slightly from the ideal 90$^\circ$, as listed in Table~\ref{lattice constant}. The TlCdS$_2$-type structure is notable as a rare example of Ag$^+$ in octahedral coordination. This rarity arises because the 4$d$ orbitals of Ag$^+$ (4$d^{10}$) are fully occupied, as are the $\sigma$-antibonding $e_{\mathrm{g}}$ orbitals in an octahedral crystal field. Consequently, Ag–ligand bonding is dominated by 5$s$/5$p$ ($sp$–$sp^3$) orbital interactions, favoring 2-4 coordination geometries (linear, trigonal planar, tetrahedral) over octahedral environments. Our first-principles calculations reveal energy differences of $-8$ and $+2$~meV/atom between the TlCdS$_2$-type and ErAgSe$_2$-type structures for ScAgSe$_2$ and TmAgTe$_2$, respectively, indicating that the TlCdS$_2$-type phase is energetically competitive for both compositions. According to Pauling’s third rule~\cite{doi:10.1021/ja01379a006}, face- and edge-sharing octahedra significantly weaken cation–anion bonds due to strong cation–cation repulsion. This effect is expected to decrease the bond strength of Ag–Se (Ag–Te) and Sc–Se (Tm–Te), which can result in intrinsically low $\kappa_{\mathrm{L}}$~\cite{https://doi.org/10.1002/adfm.202108532}.

	\vspace{0.3cm}
	\noindent \textbf{Electronic Structures.}
	The first Brillouin zone of the $P\bar{3}m1$ space group is presented in Figure~\ref{band}(b). Given the similarity in the band structures of ScAgSe$_2$ and TmAgTe$_2$, the orbital-projected band structures of ScAgSe$_2$ are shown in Figures~\ref{band}(c)–\ref{band}(j), whereas the band structure of TmAgTe$_2$ is provided in Figure~\textcolor{red}{S1}. Both ScAgSe$_2$ and TmAgTe$_2$ are identified as indirect bandgap semiconductors, with the VBM located at the A point of the first Brillouin zone and the conduction band minimum (CBM) positioned at the L point. It is noteworthy that the secondary and tertiary VBMs, situated along the $\Lambda$ (the $\Gamma$–K path) and $\Sigma$ (the $\Gamma$–M path) directions, respectively, exhibit a valley degeneracy of six. The energy differences between these local VBMs and the primary VBM are denoted as $\Delta{E}$ and $\Delta{E^{\prime}}$, respectively. As reported in Table~\ref{lattice constant}, the values of $\Delta{E}$ ($\Delta{E^{\prime}}$) are 205\,(316) meV for ScAgSe$_2$ and 165\,(246) meV for TmAgTe$_2$. The electronic structure of the TlCdS$_2$-type TmAgTe$_2$ phase differs markedly from that of the tetragonal counterpart, which features the VBM at the $\Gamma$ point~\cite{C8TA06470A}. This distinction arises from the tetrahedral coordination of Ag with Te in the tetragonal phase, enabling $p$–$d$ orbital coupling.

	Symmetry matching is one of the key prerequisites for orbital coupling in molecular orbital (MO) theory~\cite{PhysRev.32.186,hund1927deutung}, which can be interpreted using group theory in combination with elementary band representations (EBRs)~\cite{TQC}. The atomic valence-electron band representations (ABRs) of the Ag-$d$ and Se-$p$ orbitals in ScAgSe$_2$ were obtained using the POS2ABR code~\cite{pos2abr}, and the corresponding EBRs are listed in Table~\ref{EBR}. The orbital-resolved contributions to the electronic structure are presented in Figure~\ref{band}. The irreducible representations (IRREPs) of the highest valence band at the $\Sigma$, $\Lambda$, A, $\Gamma$, and L points, calculated using IRVSP~\cite{irvsp}, are indicated in Figure~\ref{band}(c). Specifically, the IRREPs at the $\Gamma$, $\Sigma$, $\Lambda$, A, and L points are $\Gamma_3^-$, $\Sigma_1$, $\Lambda_2$, A$_3^-$, and L$_2^-$, respectively. The Wyckoff positions of Sc$^{3+}$/Tm$^{3+}$, Ag$^{+}$, and Se$^{2-}$/Te$^{2-}$ correspond to 1$a$, 1$b$, and 2$d$, respectively. As the valence band is primarily derived from Ag-$d$ and Se-$p$ orbitals, only the EBRs associated with these states are reported in Table~\ref{EBR}. At the $\Gamma$ point, the IRREP of the highest valence band is $\Gamma_3^-$, which is contained exclusively in the EBR $\Gamma_3^+\oplus\Gamma_3^-$ induced by the site E@2$d$. The basis functions of E@2$d$ are $p_x$ and $p_y$, implying that only the Se-$p_x$ and Se-$p_y$ orbitals contribute to this band (see Figures~\ref{band}(h) and \ref{band}(i)), while Se-$p_z$ and all Ag-$d$ orbitals have negligible contribution. This symmetry restriction results in forbidden $p$–$d$ coupling at the $\Gamma$ point. In contrast, at the A point, the IRREP of the highest valence band is A$_3^-$, which is a common element of the EBR A$_3^-$ induced by Ag-$d$ orbitals and the EBR A$_3^+\oplus A_3^-$ induced by Se-$p$ orbitals. Therefore, $p$–$d$ orbital coupling is symmetry-allowed at A. Since the valence band originates mainly from $p$–$d^*$ antibonding states, the absence of $p$–$d$ coupling at $\Gamma$ and the presence of such coupling at A yield a higher band energy at A relative to $\Gamma$~\cite{xiong2025forbidden}, producing strong dispersion along the $\Gamma$–A direction. The second highest valence band ($\Lambda_2$) arises from allowed $p$–$d$ couplings between four Ag-$d$ orbitals ($d_{x^2-y^2}$, $d_{xy}$, $d_{xz}$, $d_{yz}$) and the three Se-$p$ orbitals, as well as from orbital–orbital interactions among these states. This accounts for its higher energy relative to the $\Gamma$ point. Similarly, the third highest valence band ($\Sigma_1$) is composed of $\Sigma_1$ states induced by Ag-$d_{z^2}$ and Se-$p_z$ orbitals, along with $\Sigma_1\oplus\Sigma_2$ states originating from Ag-$d$ ($d_{x^2-y^2}$, $d_{xy}$, $d_{xz}$, $d_{yz}$) and Se-$p_x$/$p_y$ orbitals. This configuration allows $p$–$d$ coupling between Se-$p_z$ and Se-$p_x$/$p_y$ with Ag-$d$ orbitals, in addition to inter-orbital coupling among them, resulting in a higher $\Sigma_1$ band energy compared to $\Gamma_3^-$.

	Based on the above analysis, we applied biaxial tensile strain within the $ab$ plane to ScAgSe$_2$ and TmAgTe$_2$ in order to raise the energies of the bands at the $\Lambda$ and $\Sigma$ points, both of which exhibit a sixfold valley degeneracy. As shown in Figure~\ref{bandstrain} and Table~\ref{lattice constant}, the values of $\Delta E$ and $\Delta E'$ for ScAgSe$_2$ decrease substantially from 205 and 316~meV to 35 and 105~meV, respectively, upon application of 3\% tensile strain. For TmAgTe$_2$, these values are reduced from 165 and 246~meV to 23 and 71~meV, respectively, under 2\% tensile strain. Consequently, the effective valley degeneracy $N_\mathrm{v}$ in $p$-type doped ScAgSe$_2$ and TmAgTe$_2$ increases significantly, thereby enhancing the density-of-states (DOS) effective mass $m^*_\mathrm{d}$ and improving the Seebeck coefficient $S$ of these materials. The DOS shown in Figures~\ref{bandstrain}(a) and \ref{bandstrain}(b) clearly demonstrates that strain leads to a notable increase in the DOS near the Fermi level. Specifically, a larger tensile strain produces a higher DOS in the vicinity of the Fermi level.

	Upon the application of biaxial tensile strain, the lattice constants $a$ and $b$ increase slightly, accompanied by the elongation of the Ag–Se/Ag–Te bond lengths and an enlargement of the Se–Ag–Se/Te–Ag–Te bond angles. These structural modifications result in a slight decrease in the integrated crystal orbital Hamilton population (iCOHP), as shown in Table~\ref{lattice constant}, indicating that the Ag–Se bonding interaction is weakened under tensile strain. Since the $p$–$d$ antibonding states at the A point are predominantly composed of Se-$p_x$ and $p_y$ orbitals, which are sensitive to the Ag–Se bond length, the band energy at the A point decreases more rapidly than at the $\Lambda$ and $\Sigma$ points. This differential band energy shift accounts for the reduction in $\Delta E$ and $\Delta E'$ under tensile strain.

	\vspace{0.2cm}
	\noindent \textbf{Electronic Transport Properties.}	
	The variations of $\sigma$, $S$, PF, and $m^*$ for ScAgSe$_2$ under $p$-type doping, with different applied strains, as functions of $n_\mathrm{h}$ at 300~K and 800~K are presented in Figure~\ref{sigma}. First, for both the in-plane directions ($a$- and $b$-axes) and the out-of-plane direction ($c$-axis), and regardless of strain application, $\sigma$ increases with increasing $n_\mathrm{h}$ but decreases with increasing $T$, while $S$ exhibits the opposite behavior. These are typical characters of semiconductors~\cite{2008Complex}. Secondly, as tensile strain increases, $\sigma$ gradually decreases whereas $S$ progressively increases. This behavior is attributed to the increase in $m_{\mathrm{DOS}}^*$ with strain: a larger $m_{\mathrm{DOS}}^*$ suppresses $\sigma$ but enhances $S$. Additionally, the increase in valley degeneracy $N_\mathrm{v}$ contributes to the enhancement of $S$. Moreover, an increased electron–phonon scattering rate (Figure~\textcolor{red}{S2}) further suppresses $\sigma$. Without applied strain, the in-plane $\sigma$ values are higher than the out-of-plane values. For instance, at $n_\mathrm{h} = 1\times10^{19}$~cm$^{-3}$, $\sigma$ in the in-plane directions is 13220~S\,m$^{-1}$ at 300~K and 2906~S\,m$^{-1}$ at 800~K, compared to 7089~S\,m$^{-1}$ and 1825~S\,m$^{-1}$ in the out-of-plane direction. Upon application of strain, $\sigma$ in the in-plane directions decreases more rapidly than in the out-of-plane direction, while $S$ shows an opposite trend. This contrasting behavior originates from the strain-induced increase in $m_{\mathrm{DOS}}^*$ primarily along the in-plane directions.

	Owing to the opposite variation trends of $\sigma$ and $S$ with respect to $T$, $n_\mathrm{h}$, and strain, the evolution of PF in ScAgSe$_2$ is nontrivial. In brief, for the in-plane directions, PF decreases with increasing strain at low $n_\mathrm{h}$, whereas it increases with strain at high $n_\mathrm{h}$. In contrast, for the out-of-plane direction, PF increases with strain across the entire calculated $n_\mathrm{h}$ range (1$\times$10$^{19}$~cm$^{-3}$ to 1$\times$10$^{21}$~cm$^{-3}$), with the enhancement being more pronounced at 300~K than at 800~K. It is noteworthy that, since $\kappa_{\mathrm{e}}$ is proportional to $\sigma$, the $\kappa_{\mathrm{e}}$ of ScAgSe$_2$ decreases with increasing strain (Figure~\textcolor{red}{S3}). The electronic transport properties of TmAgTe$_2$ under $p$-type doping with different applied strains are shown in Figures~\textcolor{red}{S4 and S5}, exhibiting trends similar to those of ScAgSe$_2$. The primary difference is that a smaller tensile strain is sufficient for band alignment in TmAgTe$_2$ due to its smaller $\Delta{E}$. In summary, $\sigma$ for both ScAgSe$_2$ and TmAgTe$_2$ decreases with increasing strain, whereas $S$ increases significantly, primarily due to the enhancement of $N_{\mathrm{v}}$. By tuning the delicate balance between $\sigma$ and $S$, PF can be improved through the application of a moderate tensile strain.

	\vspace{0.2cm}
	\noindent \textbf{Phonon Dispersion and Lattice Thermal Conductivity.}
	Figure~\ref{phonon} shows the phonon dispersion and phonon density of states (PhDOS) of ScAgSe$_2$ and TmAgTe$_2$ under different strains at 300 K. Since the primitive unit cells of ScAgSe$_2$ (TmAgTe$_2$) contain 4 atoms, there are a total of 12 phonon branches for each compound, including 3 acoustic and 9 optical phonon branches. The phonon frequency ($\omega$) is proportional to bonding strength ($k$) and inversely proportional to atomic mass ($M$), with $\omega \propto \sqrt{k/M}$. Therefore, heavier elements generally contribute more to low-frequency phonons, while the lighter elements mainly appear in the high-frequency region. The maxima phonon frequency of TmAgTe$_2$ is much lower than that of ScAgSe$_2$ due to the heavier $M$ of Tm and Te atoms and lower $k$ of Ag-Te chemical bond caused by the lower electronegativity of Te. Note that the sharp peaks in the PhDOS of these two compounds in the low-frequency region, which correspond to the flat phonon bands, are mainly from Ag atom even in the case of TmAgTe$_2$, though Ag has the smallest $M$ in this compound. This is because of the filling of the $p-d^*$ anti-bonding states in these two compounds~\cite{https://doi.org/10.1002/adfm.202108532}. When biaxial strain is applied, the maxima phonon dispersion frequency decreases, due to the weakening of Ag-Se bond, as discussed above. For example, the maxima phonon frequency of ScAgSe$_2$ decreases from 8.40 THz (Figure~\ref{phonon}(a)) when strain is zero to 8.09 THz (Figure~\ref{phonon}(b)) at 2\% strain, and further to 7.94 THz (Figure~\ref{phonon}(c)) at 3\% strain. The reduction in phonon frequency results in a decrease in the $\overline{\nu_\mathrm{g}}$ (Table~\ref{lattice constant}), and a smaller $\overline{\nu_\mathrm{g}}$ tends to lead to a lower $\kappa_{\mathrm{L}}$.

	\begin{figure*}[th!]
		\centering
		\includegraphics[width=1.0\linewidth]{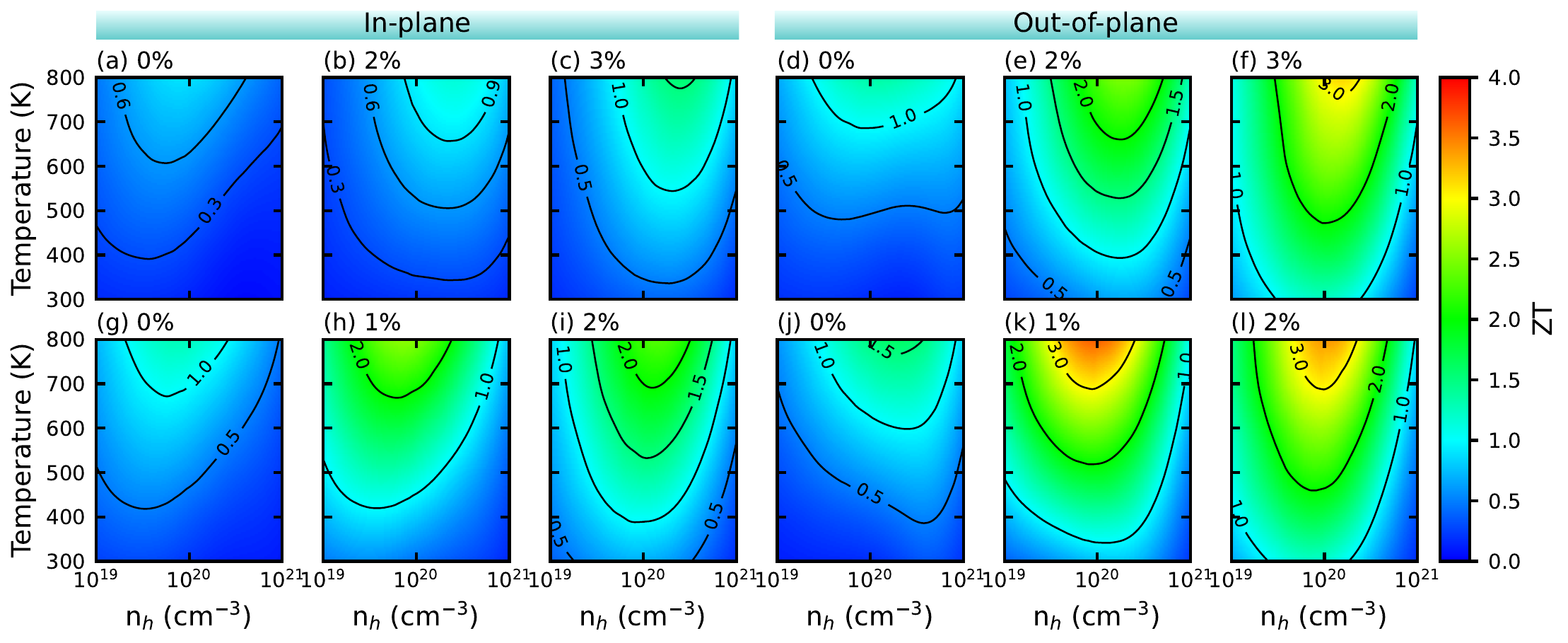}
		\caption{The $ZT^{\perp}$ and $ZT^{\parallel}$ of ScAgSe$_2$ and TmAgTe$_2$ as functions of $T$ and $n_\mathrm{h}$ under different biaxial tensile strains. (a)–(c) show the $ZT^{\perp}$ of ScAgSe$_2$ at 0\%, 2\%, and 3\% biaxial strain, respectively. (d)–(f) show the $ZT^{\parallel}$ $ZT$ of ScAgSe$_2$ at 0\%, 2\%, and 3\% biaxial strain, respectively. (g)–(i) show the $ZT^{\perp}$ of TmAgTe$_2$ at 0\%, 1\%, and 2\% biaxial strain, respectively. (j)–(l) show the $ZT^{\parallel}$ of TmAgTe$_2$ at 0\%, 1\%, and 2\% biaxial strain, respectively.}
		\label{zt}
	\end{figure*}

	The calculated $\kappa_{\mathrm{L}}$ of ScAgSe$_2$ and TmAgTe$_2$, which incorporate three-phonon ($\kappa_{\mathrm{3ph}}$) and four-phonon ($\kappa_{\mathrm{4ph}}$) scattering as well as the coherent contribution ($\kappa_{\mathrm{L}}^{\mathrm{C}}$) based on renormalized second-order force constants at each $T$, are presented in Figure~\ref{kappa}. For TmAgTe$_2$, our calculated $\kappa_{\mathrm{L}}$ at 300~K (0.71~W\,m$^{-1}$K$^{-1}$ along the $a$ axis, $\kappa_{\mathrm{L}}^{\perp}$, and 0.62~W\,m$^{-1}$K$^{-1}$ along the $c$ axis, $\kappa_{\mathrm{L}}^{\parallel}$) closely matches the reported experimental value of 0.52~W\,m$^{-1}$K$^{-1}$~\cite{tm}. For ScAgSe$_2$ at 300~K, our values ($\kappa_{\mathrm{L}}^{\perp}$ = 0.82~W\,m$^{-1}$K$^{-1}$, $\kappa_{\mathrm{L}}^{\parallel}$ = 0.55~W\,m$^{-1}$K$^{-1}$) are slightly higher than those from previous calculations~\cite{sc}, primarily because the earlier work did not include $\kappa_{\mathrm{L}}^{\mathrm{C}}$. In our study, $\kappa_{\mathrm{L}}^{\mathrm{C}}$ accounts for approximately 20\% and 30\% of the total $\kappa_{\mathrm{L}}$ for ScAgSe$_2$ and TmAgTe$_2$ at 300~K (Figure~\textcolor{red}{S6}), attributable to the strong anharmonicity of this material. ScAgSe$_2$ exhibits pronounced anisotropy in $\kappa_{\mathrm{L}}$, with $\kappa_{\mathrm{L}}^{\perp}/\kappa_{\mathrm{L}}^{\parallel} \approx 2.18$ at 300~K, whereas TmAgTe$_2$ displays much weaker anisotropy ($\kappa_{\mathrm{L}}^{\perp}/\kappa_{\mathrm{L}}^{\parallel} \approx 1.16$ at 300~K). This difference may originate from the distinct interlayer interaction strengths in the two compounds. Overall, both ScAgSe$_2$ and TmAgTe$_2$ exhibit unexpectedly low $\kappa_{\mathrm{L}}$ values relative to their average atomic masses. For example, the $\kappa_{\mathrm{L}}$ of both materials at 300~K is substantially lower than that of classical thermoelectric materials PbSe (1.6~W\,m$^{-1}$K$^{-1}$~\cite{el1983thermophysical}) and PbTe (2.0~W\,m$^{-1}$K$^{-1}$~\cite{el1983thermophysical}), even though the average atomic masses of ScAgSe$_2$ (77.69~amu) and TmAgTe$_2$ (133.00~amu) are considerably smaller than those of PbSe (143.08~amu) and PbTe (167.40~amu). The $\kappa_{\mathrm{L}}$ of TmAgTe$_2$ is lower than that of ScAgSe$_2$, primarily due to its smaller average phonon group velocity $\overline{\nu_\mathrm{g}}$ (1823~m\,s$^{-1}$, Table~\ref{lattice constant}) compared to ScAgSe$_2$ (2193~m\,s$^{-1}$). This reduction arises from the heavier atomic constituents (Tm and Te) and the weaker Ag–Te and Tm–Te bond strengths in TmAgTe$_2$. As discussed previously~\cite{https://doi.org/10.1002/adfm.202108532}, the presence of filled anti-bonding states and edge- and face-sharing octahedra contribute to the weak chemical bonds in both compounds, facilitating strong phonon–phonon scattering. As shown in Figure~\ref{ss}, both ScAgSe$_2$ and TmAgTe$_2$ exhibit pronounced three-phonon scattering in the low-frequency (1–2~THz) range, corresponding to their flat phonon dispersion branches.

	When biaxial tensile strain is applied within the $ab$ plane, both $\kappa_{\mathrm{L}}^{\perp}$ and $\kappa_{\mathrm{L}}^{\parallel}$ of ScAgSe$_2$ decrease slightly with increasing strain, consistent with the previously observed trend that the average phonon group velocity $\overline{\nu_\mathrm{g}}$ decreases with strain (Table~\ref{lattice constant}). Simultaneously, the three-phonon and four-phonon scattering rates (1/$\tau$) increase slightly with strain, further contributing to the reduction in $\kappa_{\mathrm{L}}$ due to the enhanced anharmonicity discussed above. In particular, under 3\% tensile strain, the reduction in $\kappa_{\mathrm{L}}$ is pronounced: $\kappa_{\mathrm{L}}^{\parallel}$ at 300~K and 800~K decreases from 0.57~W\,m$^{-1}$K$^{-1}$ and 0.38~W\,m$^{-1}$K$^{-1}$ (zero strain) to 0.36~W\,m$^{-1}$K$^{-1}$ and 0.27~W\,m$^{-1}$K$^{-1}$, respectively. This significant drop is mainly attributed to the relatively large four-phonon scattering rate (1/$\tau$) (Figure~\ref{ss}(b)). In contrast, $\kappa_{\mathrm{L}}^{\parallel}$ of TmAgTe$_2$ exhibits a nonmonotonic dependence on strain, first decreasing and then increasing with further strain. Specifically, at 1\% tensile strain, $\kappa_{\mathrm{L}}^{\parallel}$ values (0.39~W\,m$^{-1}$K$^{-1}$ at 300~K and 0.24~W\,m$^{-1}$K$^{-1}$ at 800~K) are lower than those at 2\% strain (0.46~W\,m$^{-1}$K$^{-1}$ at 300~K and 0.32~W\,m$^{-1}$K$^{-1}$ at 800~K). This deviates from the expected trend based on $\overline{\nu_\mathrm{g}}$, which consistently decreases with increasing strain. The anomaly arises from the unusually high four-phonon scattering rate under 1\% strain (Figure~\ref{ss}(d)), originating from its large weighted phase space (Figure~\textcolor{red}{S7}).

	\vspace{0.2cm}
	\noindent \textbf{Figure of Merit $ZT$.}
	Based on the computed electrical and phonon transport properties, the estimated dimensionless figure of merit ($ZT$) values for ScAgSe$_2$ and TmAgTe$_2$ are presented in Figure~\ref{zt}. For all cases, $ZT$ exhibits a characteristic trend of first increasing and then decreasing with $n_{\mathrm{h}}$. The maximum $ZT$ ($ZT_\mathrm{max}$) for both compounds occurs at 800~K, the highest temperature considered in this study, owing to the proportionality of $ZT$ to $T$ and the inverse correlation of $\kappa_{\mathrm{L}}$ with $T$. For the unstrained ScAgSe$_2$, the in-plane ($ZT_\mathrm{max}^{\perp}$) and out-of-plane ($ZT_\mathrm{max}^{\parallel}$) values at 800~K are 0.89 and 1.35, respectively. Under 3\% biaxial tensile strain, $ZT_\mathrm{max}^{\perp}$ and $ZT_\mathrm{max}^{\parallel}$ increase to 1.54 and 3.10, corresponding to enhancements of 73\% and 130\%, respectively. At 300~K, the application of 3\% strain increases $ZT_\mathrm{max}^{\perp}$ from 0.25 to 0.41 and $ZT_\mathrm{max}^{\parallel}$ from 0.24 to 1.23, representing percentage increases of 64\% and 412\%, respectively. For TmAgTe$_2$, $ZT_\mathrm{max}^{\perp}$ and $ZT_\mathrm{max}^{\parallel}$ at 800~K are 1.28 and 1.60, respectively, slightly exceeding those of ScAgSe$_2$ due to its lower $\kappa_{\mathrm{L}}$ and higher PF. At 800~K, 1\% tensile strain increases $ZT_\mathrm{max}^{\perp}$ and $ZT_\mathrm{max}^{\parallel}$ by 99\% and 126\%, reaching exceptionally high values of 2.55 and 3.62, respectively. At 300~K, a 2\% tensile strain enhances $ZT_\mathrm{max}^{\perp}$ and $ZT_\mathrm{max}^{\parallel}$ by 86\% and 258\%, respectively. These substantial improvements arise from the concurrent increase in PF and reduction in $\kappa_{\mathrm{L}}$ achieved through tensile strain.
	

	\vspace{0.5cm}
	\noindent $\blacksquare$ \textbf{\textcolor{darkblue}{CONCLUSIONS}} \\
    This study reveals a new mechanism for the synergistic enhancement of the PF and suppression of $\kappa_{\mathrm{L}}$ in TlCdS$_2$-type silver chalcogenides ScAgSe$_2$ and TmAgTe$_2$ under biaxial strain, based on first-principles calculations. The forbidden $p$–$d$ orbital coupling at the $\Gamma$ point and allowed $p$–$d$ coupling along the A point, $\Delta$, and $\Sigma$ lines result in high electronic band dispersion along $\Gamma$–A, $\Gamma$–K, and $\Gamma$–M. A small tensile biaxial strain elongates Ag–$X$ bonds and reduces the coupling between Se/Te-$p$ and Ag-$d$ orbitals, which aligns the valence band valleys at the A and $\Lambda$ points, increasing valley degeneracy $N_\mathrm{v}$ from 1 to 7. At the optimal hole concentration $n_\mathrm{h}$ (corresponding to $ZT_{\mathrm{max}}$), the maximum PF of ScAgSe$_2$ under 3\% strain and TmAgTe$_2$ under 2\% strain in the out-of-plane direction are enhanced by factors of 4.8 (1.4) and 1.4 (1.63) at 300~K (800~K), respectively. The filled $p$–$d$ anti-bonding states, together with the edge- and face-sharing AgSe$_6$/AgTe$_6$ octahedra, significantly weaken Ag–Se/Ag–Te chemical bonds, leading to intrinsically low $\kappa_{\mathrm{L}}$. Application of small tensile strain further weakens these bonds, reducing phonon group velocities and enhancing three- and four-phonon scattering, thereby further lowering $\kappa_{\mathrm{L}}$. For example, the $\kappa_{\mathrm{L}}$ of ScAgSe$_2$ along the $c$-axis decreases from 0.57~W\,m$^{-1}$K$^{-1}$ (0.38~W\,m$^{-1}$K$^{-1}$) to 0.36~W\,m$^{-1}$K$^{-1}$ (0.27~W\,m$^{-1}$K$^{-1}$) at 300~K (800~K) under 3\% tensile strain. Similarly, the $\kappa_{\mathrm{L}}$ of TmAgTe$_2$ along the $c$-axis decreases from 0.61~W\,m$^{-1}$K$^{-1}$ (0.42~W\,m$^{-1}$K$^{-1}$) to 0.39~W\,m$^{-1}$K$^{-1}$ (0.24~W\,m$^{-1}$K$^{-1}$) at 300~K (800~K) under 1\% tensile strain. Benefiting from the simultaneous PF enhancement and $\kappa_{\mathrm{L}}$ reduction, under 3\% biaxial strain, the maximum $ZT$ of ScAgSe$_2$ along the $c$-axis reaches 1.23 (300~K) and 3.10 (800~K), representing increases of 412\% and 130\% over the unstrained values, respectively. For TmAgTe$_2$, even higher $ZT$ values are achieved under 1\% tensile strain, with $ZT_{\mathrm{max}}$ along the $c$-axis ($a$-axis) reaching 0.79 (0.73) and 3.62 (2.55) at 300~K (800~K), corresponding to enhancements of 132\% (94\%) and 126\% (99\%), respectively. The significant $ZT$ improvements at room and intermediate temperatures are particularly important, as the average $ZT$ over the practical operating temperature range is more relevant to thermoelectric conversion efficiency than peak values. The strain-engineering strategy proposed here—rooted in molecular orbital analysis—is extendable to other materials systems and offers a general approach for boosting thermoelectric performance by concurrently optimizing electronic and phonon transport properties.

	\vspace{0.5cm}
	\noindent $\blacksquare$ \textbf{Supporting information} \\

	\noindent The Supporting Information is available free of charge on the ACS Publications website at DOI: \\

	\hspace{0.2cm}

	\noindent \textbf{Notes} \\
	The authors declare no competing financial interest.\\

	\noindent $\blacksquare$ \textbf{\textcolor{darkblue}{ACKNOWLEDGMENTS}} \\
	The authors acknowledge the support of the National Science Foundation of China (Grant No. 12374024) and Fundamental Research Funds for the Central Universities (No. FRF-BRB-25-006). The computing resource was supported by USTB MatCom of Beijing Advanced Innovation Center for Materials Genome Engineering.

	\bibliography{ref}
	
	\begin{tocentry}
		\includegraphics[clip,width=1.0\linewidth]{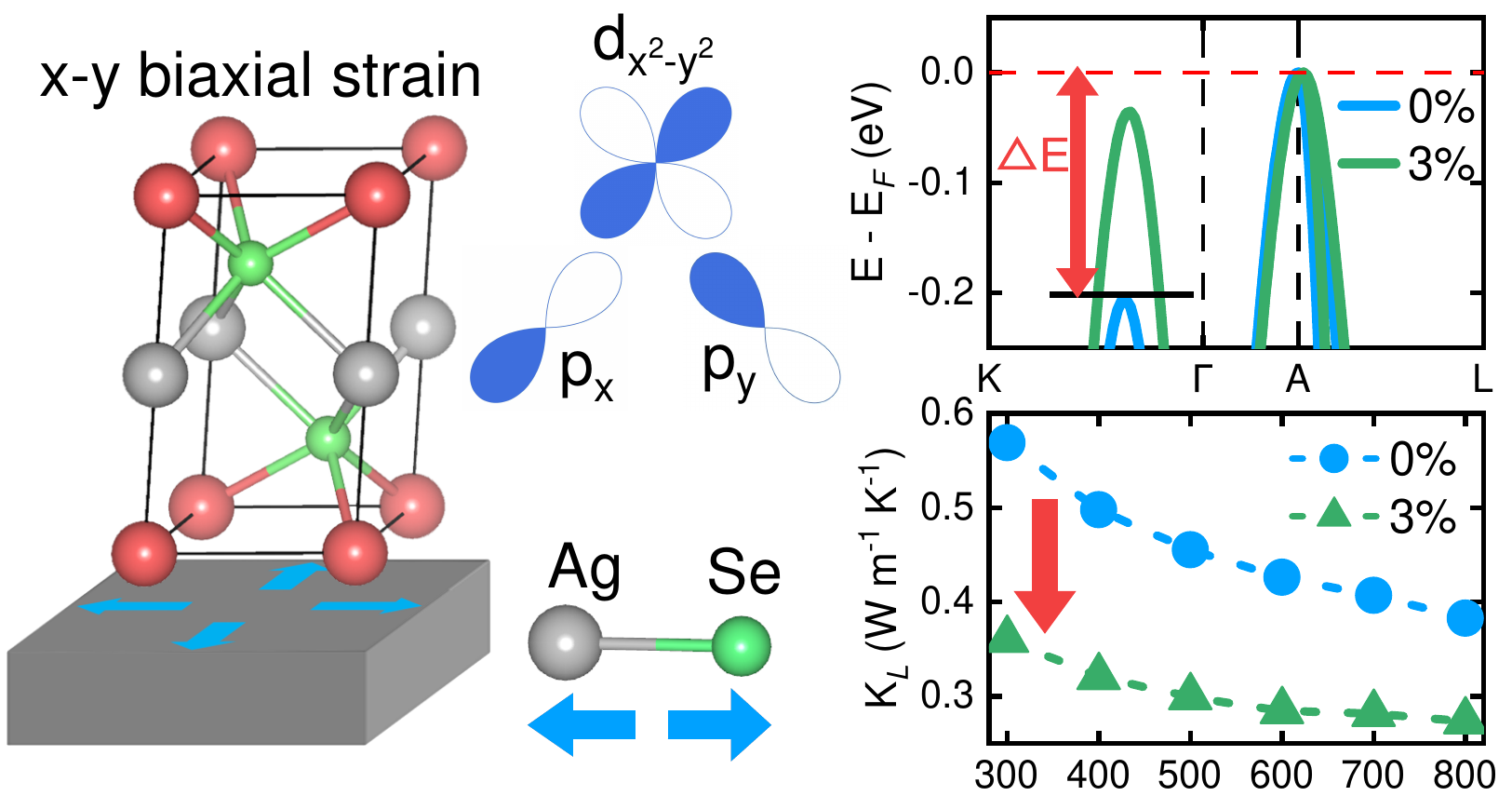}
	\end{tocentry}
	
\end{document}